\documentstyle[12pt]{article}
\newcommand{\be}{\begin{equation}}
\newcommand{\ee}{\end{equation}}
\newcommand{\bea}{\begin{eqnarray}}
\newcommand{\eea}{\end{eqnarray}}
\newcommand{\al}{\alpha}

\newcommand{\dl}{\delta}
\newcommand{\Dl}{\Delta}

\newcommand{\et}{\eta}
\newcommand{\th}{\theta}
\newcommand{\Th}{\Theta}
\newcommand{\thv}{\vartheta}

\newcommand{\lm}{\lambda}
\newcommand{\ks}{\xi}

\newcommand{\om}{\omega}
\newcommand{\Om}{\Omega}

\newcommand{\rarrow}{\rightarrow}
\newcommand{\nn}{\nonumber}

\begin{document}

\title{Kinetic description of particle interaction
with a gravitational wave}

\author{A. Anastasiadis, K. Kleidis and H. Varvoglis \\
{\small Section of Astrophysics, Astronomy and Mechanics} \\
{\small Department of Physics,} \\
{\small Aristotle University of Thessaloniki,} \\ 
{\small GR-54006 Thessaloniki, Greece } }

\maketitle

\begin{abstract}

The interaction of charged particles, moving in a uniform 
magnetic field, with a plane polarized gravitational wave 
is considered using the Fokker - Planck - Kolmogorov (FPK) 
approach. By using a stochasticity criterion, we determine 
the exact locations in phase space, where resonance overlapping 
occurs. We investigate the diffusion of orbits around each 
primary resonance of order $m$ by deriving general analytical 
expressions for an effective diffusion coefficient. A solution 
of the corresponding diffusion equation (Fokker-Planck equation)
for the static case is found. Numerical integration of 
the full equations of motion and subsequent calculation of the 
diffusion coefficient verifies the analytical results.  

\end{abstract}

\section{INTRODUCTION}

Many efforts that have been made to detect gravitational waves gave, so far,
no convincing evidence that they were actually being seen [1]. This is due 
to the fact that not only their amplitude is very small [2], but it is highly 
possible that some kind of damping mechanism operates on them as they travel 
through space [3-5]. This damping may originate in the interaction of the 
gravitational wave with the interstellar matter [6,7].

In a recent paper [8], which hereafter is referred to as Paper I, the problem 
of the interaction of a charged particle with a gravitational wave,
in the presence of a uniform magnetic field, has been considered for various 
directions of propagation of the wave with respect to the magnetic field. It 
was found that, in the oblique propagation case the motion of the particle 
becomes chaotic and may be considered as a diffusion in momentum space, 
provided that its initial momentum is sufficiently large. 

In order to address in detail the interaction of charged particles with a 
gravitational wave, one should try to calculate the diffusion rate (in
momentum space) of the particles which follow chaotic trajectories. This task
involves the derivation of a Fokker - Planck (FP) type diffusion equation 
and the calculation of the corresponding diffusion coefficient [9].

In the present paper we investigate the energy diffusion of charged particles 
in the presence of a uniform magnetic field, $\vec{B}=B_{0}{\hat e_{\rm z}}$, 
due to their non-linear interaction with a linear polarized gravitational 
wave, propagating obliquely with respect to the direction of the magnetic 
field ($20^o\le \th \le 60^o$). The analysis is carried out in the framework 
of the weak field theory, considering the gravitational wave as a small 
pertubation in a flat space time. We use the Fokker - Planck - Kolmogorov 
(FPK) approach and refer to the globally stochastic regime, where overlapping 
of many resonances occurs. In a partially stochastic regime the FKP approach 
can not be applied, as the particles do not undergo "normal diffusion" 
(random walk process) but rather follow Levy statistics [10]. This statistical 
approach is possible only after deriving general formulas, that hold for every 
value of the perpendicular energy of the charged particle and not just for the 
highest ones, as it has been considered in Paper I.

The motion of a charged particle in curved spacetime is given, in Hamiltonian 
formalism [11], by the differential equations
\be
{{\rm d} x^{\mu} \over {\rm d} {\lm}} = {\partial H \over \partial {\pi}_{\mu}} \; , 
\; \; \; {{\rm d} {\pi}_{\mu} \over {\rm d} {\lm}} = -{\partial H \over \partial x^{\mu}}
\ee
where ${\pi}_{\mu}$ are the generalized momenta (corresponding to the 
coordinates $x^{\mu}$) and the "super Hamiltonian", $H$, is given by 
the relation
\be
H={1 \over 2} g^{\mu \nu} \left ( {\pi}_{\mu} - e A_{\mu} \right ) \left (
{\pi}_{\nu} - e A_{\nu} \right ) \equiv {1 \over 2}
\ee
(in a system of geometrical units where $\hbar = c = G = 1$) In Eq.(2) 
$g^{\mu \nu}$
denotes the components of the contravariant metric tensor, which are 
defined as
\be
g^{\mu \nu} = {\et}^{\mu \nu} + h^{\mu \nu}
\ee
with ${\et}^{\mu \nu} = diag (1, -1, -1, -1)$ and $ \vert h^{\mu \nu} 
\vert \ll 1$. $A_{\mu}$ is the vector potential, corresponding to the 
tensor of the electromagnetic field in a curved spacetime $F_{\mu \nu}$. 
The mass of the particle is taken equal to 1. For the specific form of 
the magnetic field we may take
\be
A^{0} = A^{1} = A^{3} = 0 \; , \; \; \;  A^{2} = B_{0} x^{1}
\ee

\section{THE STOCHASTICITY CRITERION}

We consider the case of a charged particle moving in the curved spacetime
background of a linear polarized gravitational wave, which propagates
obliquely with respect to the direction of a uniform and static, in time,
magnetic field, $\vec{B} = B_{0}{\hat e_{\rm z}}$. The non-zero components of
the metric tensor are presented in Paper I (see references therein) 
and we normalize lengths and time to ${c \over \Om}$, where
$\Om$ is the Larmor angular frequency. We, furthermore, eliminate 
one degree of freedom from our dynamical system through the 
canonical transformation
\bea
\thv^3  & = &  x^0 \; - \; cos \th \: x^3 ,\; \; \pi_3 \; = \; - \: cos \th 
\: I_3 \nn \\
\thv^0 \; & = & \; x^0 , \; \; \pi_0 \; = \; I_3 \; + \; I_0
\eea
Accordingly, the problem of the motion of a charged particle in a gravitational 
wave is reduced to a two-degrees of freedom dynamical system [8], and the 
"super Hamiltonian" (2), in this case, is written in the form
\bea
H & =& {1 \over 2} I_3^2 - {1 \over 2} {1 + \al sin^{2} \th 
sin ( \nu \Th ) \over 1 + \al sin ( \nu \Th )} {\pi}_{1}^{2} \nn
\\
&-& {1 \over 2} { ( x^{1} )^{2} \over 1 - \al sin ( \nu \Th)} - {1 \over 2} 
{1 + \al cos^{2} \th sin ( \nu \Th ) \over 1 + \al sin ( \nu \Th )}
cos^2 \th I_{3}^{2}\nn  
\\
&+& {1 \over 2} {\al sin 2 \th cos \th sin ( \nu \Th ) \over 1 + \al sin 
\nu \Th )} {\pi}_{1} I_{3}
\eea
where we have set 
\be
\Th \; = \; sin \th \: x^1 - \thv^3
\ee
In Eq. (6) $\al$ is the normalized, dimensionless amplitude of the 
gravitational wave and $\nu = {\om \over \Om}$ denotes the dimensionless 
frequency.

The dynamical system under consideration possesses chaotic regions in phase 
space when $\al \neq 0$ [7,8,12]. In order to examine the transition from 
regular to stochastic motion we use {\em Chirikov's overlap criterion} [13,14] 
to obtain the lowest amplitude of the gravitational wave, ${\al}_{\rm thr}$, 
above which the dynamical system shows prominent chaotic behaviour. 

We first write the Hamiltonian (6) in {\em action-angle} variables through 
the canonical transformation 
\be
x^1 \; = \; (2I_1)^{1/2} sin \thv^1, \; \; \pi_1 \; = \; (2I_1)^{1/2} cos \thv^1,
\ee
and, since $\al \ll 1$, we make the approximation
\be
{1 \over 1 \pm \al sin ( \nu \Th )}  \approx  1 \mp \al sin ( \nu \Th )
\ee
The resulting Hamiltonian is of the form
\be
H = H_{0} + \al H_{1} sin ( \nu \Th )
\ee
We expand the trigonometric term of the perturbation in a Fourier series [7]. 
After further manipulation, the Hamiltonian (10) is written in the form
\bea
H&=&{1 \over 2} sin^{2} \th I_{3}^{2} - I_{1} \nn \\
&+&{ \al \over 2} \left [ - I_{1} sin^{2} \th {\sum_{\ell = - \infty}^{\infty}} 
J_{\ell} ( \nu r ) sin ( \ell {\thv}^{1} - \nu {\thv}^{3} ) \right . \nn \\
&+&I_{1} ( 1 + cos^{2} \th ) {\sum_{\ell = - \infty}^{\infty}} [ 
2 J_{\ell}^{\prime \prime} ( \nu r ) + J_{\ell} ( \nu r ) ] 
sin ( \ell {\thv}^{1} - \nu {\thv}^{3} ) \nn \\
&+&cos^{2} \th sin^{2} \th I_{3}^{2} {\sum_{\ell = - \infty}^{\infty}} J_{\ell} 
( \nu r ) sin ( \ell {\thv}^{1} - \nu {\thv}^{3} ) \nn \\
&+& \left . I_{3} sin 2 \th cos \th {\sum_{\ell = - \infty}^{\infty}} 
{\ell \over \nu} 
J_{\ell} ( \nu r ) sin ( \ell {\thv}^{1} - \nu {\thv}^{3} ) \right ]
\eea
where $J_{\ell} ( \ks )$ is the Bessel function of order $\ell$, 
$r = (2I_{1})^{1/2} sin \th $ is the linear momentum along the x-axis 
and a prime denotes differentiation with respect to $\ks = \nu r$. The 
perturbation term of the Hamiltonian function, $H_1$, depends on an 
infinite series of linear combinations of the angles $\thv^{1}$ 
and $\thv^{3}$, a fact that leads to resonances. In this case, Chirikov's 
criterion states that chaos appears when the width of a resonance, $\dl I_{1}$, 
becomes larger or equal than the distance between two consecutive first order
resonances, $\Dl I_{1}$.

By a near identity transformation we remove all trigonometric terms from
$H_1$, except from the one of order $\ell = m$, which generates the principal 
resonance and corresponds to the family of islands whose width enters in 
the stochasticity criterion [7]. The resulting Hamiltonian contains only the 
integrable part $H_0$ and the dominant term and it is therefore called 
{\em resonant Hamiltonian}, $H_{\rm R}$ [15,16]. Performing the canonical 
transformation
\bea
{\thv}^{1 \ast } & = & {\thv}^{1} - { \nu \over m} {\thv}^{3} \; \; , \; \;
I_{1}^{ \ast } = I_{1} \nn \\ 
{\thv}^{3 \ast } & = & {\thv}^{3} \; \; , \; \; \;  I_{3}^{ \ast } = I_{3} + 
{ \nu \over m} I_{1}
\eea
the resonant Hamiltonian is finally written in the form
\bea
H_{\rm R}&=&{1 \over 2} sin^{2} \th ( I_{3}^{ \ast} - { \nu \over m} I_{1} )^{2} 
- I_{1} \nn \\
&-&{\al \over 2} \left [ I_{1} sin^{2} \th J_{m} ( \ks ) - I_{1} (1 + 
cos^{2} \th ) [ 2 J_{m}^{\prime \prime} ( \ks ) + J_{m} ( \ks ) ] \right . \nn \\
&-&cos^{2} \th sin^{2} \th (I_{3}^{ \ast} - { \nu \over m} I_{1} )^{2} 
J_{m} ( \ks ) \nn \\
&+&2 \left . (I_{3}^{ \ast} - {\nu \over m} I_{1} ) cos^{2} \th 
{m \over \nu} J_{m} ( \ks ) \right ] sin ( m {\thv}^{1 \ast} )
\eea
Since $ {\thv}^{3}$ is a cyclic coordinate, the corresponding generalized 
momentum $I_{3}^{ \ast}$ will be a constant of the motion, so that the 
dynamical system becomes one-degree of freedom. Hamiltonian (13) describes 
the motion of a particle around each first order resonance. 
Using the resonant condition
\be
m {{\rm d} {\thv}^{1} \over {\rm d} \lm } = \nu {{\rm d} {\thv}^{3} \over {\rm d} \lm }
\ee
and the fact that $H_{0} \approx {1 \over 2}$, we find the order $m$ of the 
dominant resonance
\be
m = \nu ( 1 + 2 I_{1} )^{1/2} sin \th
\ee
In this case, $ m \neq \nu r $ and not $m \simeq \nu r$, which was the case 
considered in Paper I, for 
$I_{1} \gg 1$. This is because in the present paper we are interested in a 
general formula for the stochasticity threshold, valid for every $I_{1}$. 
The distance $\Dl I_{1}$, between two consecutive first order resonances 
is calculated by Eq.(13) and the fact that $\Dl m = 1$,
\be
\Dl I_{1} = {( 1 + 2 I_1 )^{1/2} \over \nu sin \th }
\ee
while the corresponding {\em resonant width} $\dl I_{1}$ is given by [15,16] 
\bea
\dl I_{1} & = & \left [ { 8 \al m^{2} \over \nu^{4} sin^{4} \th }  \left \vert 
( m^{2} - \nu^{2} sin^{2} \th ) (1 + cos^{2} \th ) J_{m}^{\prime \prime} \right . 
\right . \nn \\
& + & \left . \left . ( 4 m^{2} - \nu^{2} sin^{2} \th ) cos^{2} \th J_m 
\right \vert \right ]^{1/2}
\eea
Then, Chirikov's criterion, $\dl I_{1} \geq \Dl I_{1}$, reads
\bea
{1 \over \al} & \leq & 8 \vert ( m^{2} - \nu^{2} sin^{2} \th ) 
(1 + cos^{2} \th ) J_{m}^{\prime \prime}  \nn \\
& + & ( 4 m^{2} - \nu^{2} sin^{2} \th ) cos^{2} \th J_m \vert 
\eea
The above relation is the most general form of the stochasticity criterion and 
holds for any value of $I_{1}$, $\nu$ and $\th$. We see that 
for $I_{1} \gg 1$ it reduces to 
\be
8 {\nu}^{2} r^{2} \vert 4 cos^{2} \th J_{m} + (1 + cos^{2} \th ) 
J_{m}^{\prime \prime} \vert \geq {1 \over \al}
\ee
which is the corresponding result of Paper I. In this approximation, 
we may obtain an asymptotic form of the stochasticity threshold, by
taking $ r \rarrow \infty $ [8]. We obtain
\be
\al \geq 0.07 {1 \over ( \nu r )^{5/3} } {1 \over cos^{2} \th} 
\ee
We see that the stochasticity threshold is a rapidly decreasing function 
of $ \nu$ and $ r $. Therefore chaotic behaviour will appear, no matter 
how small the amplitude of the gravitational wave might be, provided that 
the initial momentum of the particles is sufficiently large. It is clear
that in this case the high order principal resonances will overlap.
On the other hand, if $I_1$ is small the chaotic behaviour will appear only
if the wave amplitude is considerably large and the relative frequency small,
leading to the overlapping of the low order resonances.

Following the above argument, in Fig.1 we give the stochasticity threshold, 
$ \al_{thr} $, as a function of the low order resonance $m$, for different 
values of the wave propagation angle $\th$ and $\nu=1.8$. Notice that 
$ \al_{thr} < 0.2 $ and decreases rapidly as the order of resonance $m$ 
increases. 

\section{THE FPK APPROACH}

\subsection{ANALYTIC RESULTS} 

Following the FPK approach [9], a diffusion equation for the energy 
distribution function of particles averaged over the phases, ${\cal F} 
( I_{1} , t )$, can be written for the system described by the Hamiltonian (13)
\be
{ \partial {\cal F} \over {\partial t}} = {1 \over 2} { \partial \over 
{\partial I_{1}}} \left ( D ( I_{1} ) { \partial {\cal F} \over 
{\partial I_{1}}} \right )
\ee
To lowest order in $ \al $, Eq. (21) describes a diffusion 
process in the variable $I_{1} = {p_{x}^{2} \over 2}$ at constant $I_{3}$.
The actual expression for the diffusion coefficient $D(I_{1})$ depends on 
the assumptions for the phase dynamics [9,17]. In the random phase 
approximation, it reduces to the quasilinear result [18,19]
\be
D ( I_{1} ) = \pi { \al }^{2} \sum_{m} m^{2} H_{1}^{2} \; \; \; \dl \left ( \: 
{{\rm d} \thv^{1 \ast} \over {\rm d} \lm} \: \right )
\ee
which, in our case, reads
\be
D ( I_{1} ) = \pi { \al }^{2} \sum_{m} m^{2} H_{1}^{2} \; \; \; \dl \left ( m^2 - 
\nu^2 sin^2 \th \: [ 1+ 2 I_{1} ] \right )
\ee
Around each principal resonance of order $m$ we may associate an {\em effective 
diffusion coefficient}, $D_m$, by averaging $D(I_{1})$ over the region between 
two successive first order resonances
\be
D_m \equiv < D (I_1) > = {1 \over \Dl m} \int_m^{m+1} D(I_1) dm
\ee
To calculate $D_m$ we use the facts that $\Dl m = 1$ and 
\be 
\dl \left ( f(m) \right ) = {\dl ( m \; - \; m_0 ) \over \vert f^{\prime}(m_0) 
\vert }
\ee
where $m_0$ is a simple zero of $f(m)$ which, in this case, is given by Eq.(15) 
[20]. Accordingly, we obtain
\be
D_{m} \equiv < D (I_{1 m}) > = {1 \over 2} \pi {\al}^{2} m H_{1 m}^{2}
\ee
where $I_{1 m}$ is the value of $I_{1}$ at each principal resonance of order $m$, 
which is found, from Eq.(15), to be of the form
\be
I_{1 m} = {1 \over 2} \left ( {m^{2} \over {\nu}^{2} sin^{2} \th} - 1 \right )
\ee
and $H_{1 m}$ corresponds to the perturbation term of the Hamiltonian (13) 
for $I_{1} = I_{1 m}$. Eq.(26), in terms of $I_{1 m}$, reads
\bea
D_{m} &=& {1 \over 2} \pi {\al}^{2} \nu sin \th (1 + 2 I_{1 m} )^{1/2} \nn \\ 
&\times& \left [ I_{1 m} ( 1 + cos^{2} \th ) J_{m}^{\prime \prime} + ( {3 \over 2} + 4 I_{1m} ) 
cos^{2} \th J_{m} \right ]^{2}
\eea
We use the above relation in order to determine the analytical values
of the diffusion coefficient, as it holds for any value of the parameters.
It is clear that the diffusion coefficient scales with the wave amplitude 
$\al$ and, through the value of $I_{1 m}$, with the order of resonance $m$. 
The diffusion coefficient reaches high values at low order resonances 
(small $I_{1m}$) when the wave amplitude is large. In the opposite case 
(small $\al$) the diffusion becomes effective in the range of high order 
resonances and, thus, in large $I_1$. In both cases for a given $\al$
the diffusion increases as the action increases.

\subsection{NUMERICAL RESULTS}

For the sake of numerical simplicity and in order to speed up the numerical 
integrations we investigate the case of low order resonances using $\al=0.2$ 
and $\nu=1.8$ throughout the whole of our numerical calculations. Since the 
results scale with the amplitude of the wave, $\al$, the calculated diffusion 
coefficient is also expected to describe, at least qualitatively, the 
diffusive acceleration at more realistic values of $\al$.

In order to verify that diffusion of particles, due to their interaction with 
the gravitational wave, occurs, we follow the orbits of a particle distribution 
on a surface of section, defined  as the surface $\nu x^0=2n\pi$. In Fig. 2 
the $\pi_1$ versus $x^1$ plot and the time variation of the action $I_1$ of 
the distribution of orbits ($N=1000$  with initial $I_1=106.5$) are presented 
for the case $\th=20^o$. 

Notice the distortion of the principal resonance (of order $m=9$) due to the 
overlapping of the secondary resonances. This is due to the fact that, for the 
parameters used, the wave amplitude is large. Thus the overlapping occurs in a 
small time-scale. The diffusion in energy, in this case, is verified from the 
considerable spread around the initial action value $I_1=106.5$.   

The numerical estimation of the diffusion coefficient is based on the 
integration of the Hamilton's equations of motion for a number of particles 
($N=1000$), having the same initial action $I_{1}$ and uniform angle 
distribution. The local diffusion coefficient is related to the average 
variations of the action $I_{1}$ through the expression [18,19]
\be
D(I_1)\simeq {< (\Delta I_1)^2 > - 2 (<\Delta I_1>)^2 \over {t}}
\ee
where
\be
<\Delta I_1>~=~\sum_{j=1}^{N} {I_{1j}(t) - I_{1j}(0) \over {N}}
\ee
and
\be
<(\Delta I_1)^2>~=~\sum_{j=1}^{N} {\left [ I_{1j}(t) - I_{1j}(0) \right ] ^2
 \over {N}}
\ee

We have performed a number of computational runs, varying the initial value
of $I_1$ $(2 \le I_1 \le 128)$ and the propagation angles $(20^o \le \theta 
\le 60^o)$. In 
Fig. 3 the numerical estimated local diffusion coefficients for $I_1=3.0$ 
and 7.0, corresponding to $m=4 $ and 6 respectively, and for $\theta =60^o$ 
are presented. Notice that the integration time is short, as for longer times 
diffusion over a large number of harmonics dominates, causing strong variations 
to the estimation of $D$. The plateau value of $D$ is chosen as the diffusion 
coefficient for the above actions. 

In Fig. 4 the analytical and the numerical effective diffusion coefficient, as 
a function of the action $I_1$, for different angles $\theta$ are shown. 
Notice that the diffusion coefficient depends strongly on the propagation angle. 
There exists a good agreement between the numerically and analytically estimated
values, indicating a power law dependence of the diffusion coefficient upon the 
action, of the general form
\be
D(I_1) \simeq d_0~I_1^k
\ee
where the values (analytically and numerically estimated) of the constant 
$d_0$ and the index $k$, with respect to the propagation angle $\theta$ are 
given in Table 1. The relative error between the analytically and numerically 
estimated values of the index $k$ varies from $7 \%$ to $24 \%$.

\section{SOLUTION TO THE DIFFUSION EQUATION}

We can easily solve the diffusion equation for the static case, i.e. 
$ \partial_{t} {\cal F} = 0$. Then Eq. (21) becomes 
\be
D(I_1) {{\rm d}^2 {\cal F} \over {\rm d}I_1^2} \: + \: {{\rm d} D(I_1) \over {\rm d}I_1} 
{{\rm d} {\cal F} \over {\rm d}I_1} \; = \; 0
\ee

We substitute the diffusion coefficient from Eq. (32) to find the solution in 
a power law form
\be
{\cal F}(I_1)~=~{{\cal F}_o \over{1-k}}~ I_1^{-(k-1)} 
\ee
with ${\cal F}_o$ constant. We must emphasize that the above solution is valid 
for relatively small values of $I_1$. In the case of very large energies 
($I_{1} \gg 1$) an analytic solution of the diffusion equation can be found, 
by considering the asymptotic form of the effective diffusion coefficient in 
the large energies approximation. For $I_{1} \gg 1$, we have $r \rarrow \infty$ 
and therefore $m \approx \ks = \nu r$. Then, the perturbation term of the 
Hamiltonian (13) reads
\be
H_{{\rm R}1} = - I_{1} \left [ ( 1 + cos^{2} \th ) J_{m}^{ \prime \prime} ( m ) + 4 
cos^{2} \th J_{m}( m ) \right ]
\ee
In this case, the Bessel equation becomes
\be
J_{m}^{\prime \prime} (m) \simeq - {J_{m}^{ \prime} (m) \over m}
\ee
and the asymptotic expansions 
\be
J_{m} (m) \sim 0.45 m^{-1/3} \; \; , \; \; \; J_{m}^{\prime} (m) \sim 0.41
m^{-2/3}
\ee
hold [21]. Therefore, in the large energies approximation, the effective 
diffusion coefficient reads
\be
D_{m} (I_{1}) = A I_{1}^{5/6} - B I_{1}^{9/6} + C I_{1}^{13/6}
\ee
where
\bea
A & = & 2^{1/2} \pi {\al}^{2} \nu sin \th \; \; a^{2} \; \; , \; \; B = 
2^{3/2} \pi {\al}^{2} \nu sin \th \; \; a b \; , \nn \\
C & = & 2^{1/2} 
{\al}^{2} \nu sin \th \; \; b^{2}
\eea
and
\be
a = {0.41 \over 2^{5/6}} {1 \over {\nu}^{5/3}} {1 + cos^{2} \th \over 
sin^{5/3} \th} \; \; , \; \; b = {1.8 \over 2^{1/6}} {1 \over {\nu}^{1/3}} 
{ cos^{2} \th \over sin^{1/3} \th }
\ee
Accordingly, the diffusion equation, in the static case, reads
\be
{\cal F} = c \int {d I_{1} \over A I_{1}^{5/6} - B I_{1}^{9/6} + C 
I_{1}^{13/6}}
\ee
where c is an integration constant. Evaluation of the integral on the r.h.s 
of Eq. (41) is possible only when $ I_1 \neq ({ a \over b})^{3/2} $, for 
which the energy distribution function appears a simple pole of order 2 [22]. 
In this case, we obtain 
\bea
{\cal F} & = & {6 c \over C} b^{2} \left [ {I_1^{1/6} \over 
(a - b I_1^{2/3}) } \right . \nn \\
& + & \left . 3 {d \over 4 a} \; \left ( \ell n { I_1^{1/6} + d \over 
I_1^{1/6} - d} \; + \; 2 tan^{-1} \: {I_1^{1/6} \over d} 
\right ) \right ]
\eea
where $d = ( {a \over b})^{1/4}$. This result is simplified considerably in 
the perpendicular propagation case, i.e. $ \th = {\pi \over 2}$, for which
Eq. (41) gives
\be
{\cal F} =  {6 c \over A} \;  \; I_1^{1/6}
\ee

\section{DISCUSSION AND CONCLUSIONS}

We have studied the interaction of a charged particle, with a plane polarized 
gravitational wave propagating obliquely ($20^o\le \th \le 60^0$) with respect 
to the direction of the ambient uniform magnetic field. 

On the basis of Hamiltonian pertubation theory, previous work on this problem
shows that the motion of the particles becomes chaotic [8]. Following this, 
we have derived analytical expressions for the stochasticity criterion, thus 
determining where, in phase space, resonance overlapping occurs, without 
any assumption regarding the values of the action and the propagation angle 
of the wave.

We have verified that diffusion of the particles in action $I_1$ occurs and
we have applied the FPK approach, in order to derive analytical general 
expressions for the effective diffusion coefficient. Numerical integration
of the exact equations of motion for particle distributions with the same 
initial action $I_1$ was also performed for the numerical estimation of the 
diffusion coefficient. 

Both methods (analytical and numerical) revealed a power law dependence of 
the diffusion coefficient upon the action $I_1$ giving similar results, with 
small variations, on the power law index. Based on these results a steady 
state solution of the Fokker - Planck diffusion equation was found.
 
The diffusion coefficient scales with the wave amplitude $\al$ and the order 
of resonance $m$ (and/or through the resonance condition with the action $I_1$).
For small $\al$ the diffusion is effective in high order resonances and thus 
in sufficient large actions. Diffusion of particles is present in low order 
resonances (small values of the action) only when the wave amplitude is large. 
In both cases the diffusion is increasing when the action is increasing. 

There is also a strong relation between the diffusion coefficient and the
propagation angle. As the angle decreases, the diffusion coefficient increases. 
This is due to the fact that the lower the angle, the greater is the amplitude 
of the wave for which stochastic motion occurs, leading to the fact that more 
resonances can overlap.  

In conclusion, we believe that the FPK approach may describe in a good 
approximation the interaction of charged particles with a gravitational wave 
in the framework of the weak field theory, where the gravitational wave is 
just a small pertubation in a flat spacetime. It is clear that more work has 
to be done in the realistic case of a curved spacetime and in the full 
non-linear theory.\\

{\bf Acknowledgements:} The authors would like to express their gratitude 
to Dr. L. Vlahos and Dr. D. Papadopoulos for their comments and advice 
during many useful discussions. One of us (K. K.) would like to thank the 
Greek State Scholarships Foundation for the financial support during this 
work. This work is partially supported by the scientific program PENED 1451 
(Greece).

\section*{REFERENCES}

\begin{itemize}

\item[1]Thorne K.S. (1987). {\em Gravitational Radiation}, in: Hawking S.W. and 
Israel W. (eds.) {\em Three Hundred Years of Gravitation}, Cambridge University 
Press, Cambridge.
\item[2]Smarr L. (1979). {\em Sourses of Gravitational Radiation}, Cambridge 
University Press, Cambridge.
\item[3]Esposito F.P. (1971). ApJ 165, 165.
\item[4]Macedo P.G. and Nelson A.H. (1983). Phys. Rev. D 28, 2382.
\item[5]Papadopoulos D. and Esposito F.P. (1985) ApJ 282, 330.
\item[6]Macedo P.G. and Nelson A.H. (1990). ApJ 362, 584.
\item[7]Varvoglis H. and Papadopoulos D. (1992). A \& A  261, 664.
\item[8]Kleidis K., Varvoglis H. and Papadopoulos D. (1993). A \& A 275, 309.
\item[9]Zaslavskii G.M. (1985). {\em Chaos in Dynamic Systems}, Harwood 
Academic, New York.
\item[10]Varvoglis H. and Anastasiadis A. (1996) AJ 111, 1718.
\item[11]Misner C.W., Thorne K.S. and Wheeler J.A. (1973). {\em Gravitation}, 
Freeman, San Francisco.
\item[12]Kleidis K., Varvoglis H., Papadopoulos D. and Esposito F. P. (1995) 
A \& A 294, 313.
\item[13]Chirikov B.V. (1969). {\em Research Conserning the Theory of 
Non-linear Resonance and Stochasticity}, Prepr. 267, Inst. Nuclear 
Physics, Novosibirsk. (Engl. Trans. CERN Transl. 31, 1971)
\item[14]Chirikov B.V. (1979). Phys. Rep. 52, 264.
\item[15]Ford J. (1978). {\em A Picture Book in Stochasticity}, in: Jorna S. 
(ed.) {\em Topics in Non-linear Dynamics}, AIP, New York.
\item[16]Greene J. (1980). Ann. N.Y. Acad. Sci. 357, 80.
\item[17]Hizanidis K., Vlahos L. and Polymilis C. (1989). Phys. Fluids B 1, 682.
\item[18]Farina D., Pozzoli R., Mannella A. and Ronzio D. (1993). Phys. Fluids 
B 5, 104.
\item[19]Farina D., Pozzoli R. and Rome M. (1994). Phys. Plasmas 1, 1871.
\item[20]Kanwal R. P. (1983). {\em Generalized Functions: Theory and Techniques}, 
Academic Press, New York.
\item[21]Abramowitz M. and Stegun A.I. (1970). {\em Handbook of Mathematical 
Functions}, Dover, New York (Eqs. 9.3.31, 9.3.33).
\item[22]Gradshteyn I.S. and Ryzhik I.M. (1965). {\em Table of Integrals, Series 
and Products}, Academic Press, New York.

\end{itemize}

\newpage

\begin{center}

\section*{TABLES}

\vspace{.5in}

\begin{tabular}{|l||l|l||l|l|} \hline

\multicolumn{1}{|c||}{$\theta$}& \multicolumn{2}{c||}{$d_0$}& \multicolumn{2}{c|}{$k$} \\ \hline
         & analytical & numerical & analytical & numerical \\ \hline
$20^o$   & $0.095$    & $0.037$   & $2.254$    &  $2.455$  \\
$35^o$   & $0.076$    & $0.091$   & $2.223$    &  $2.153$  \\ 
$45^o$   & $0.052$    & $0.025$   & $2.160$    &  $2.407$  \\
$60^o$   & $0.017$    & $0.020$   & $2.037$    &  $1.980$  \\ \hline

\end{tabular}

\vspace{.5in}

{\Large Table Caption} 

\vspace{.25in}

{\bf Table I:} The analytically and numerically estimated values of $d_0$ 
and the index $k$ with respect to the propagation angle $\theta$.

\end{center}

\newpage

\section*{FIGURE CAPTIONS}

{\bf Figure 1:} The stochasticity threshold $ \al_{thr} $
versus the order of resonance $m$ for different values of the wave 
propagation angle $\theta$ and $\nu=1.8$ \\

{\bf Figures 2a and 2b:} Surface of section plots for the case $\al =0.2$,
$\nu =1.8$, $\th=20^o$. $N=1000$ orbits with initial $I_1=106.5$
are presented: {\bf (a)} The 
$\pi_1$ versus $x^1$ plot. {\bf (b)} The time variation of the 
action $I_1$ of the distribution of orbits.\\

{\bf Figure 3:} The numerical estimated
local diffusion coefficients for $I_1=3.0$ and 7.0, corresponding to $m=4 $
and 6 respectively, for $\theta =60^o$ and $\nu=1.8$ \\

{\bf Figures 4a and b:} The diffusion coefficient 
$D(I_1)$, as a function of the perpendicular energy of the charged particle, 
$I_1$ for different angles of propagation, with  $\al =0.2$ and $\nu= 1.8$. 
The $\Box$ are analytical values and the $\triangle$ are numerical ones: 
{\bf (a)} For $\th = 20^o$ and $35^o$. {\bf (b)} For $\th = 45^o$ and $60^o$. \\

\end{document}